\documentstyle[a4,12pt,epsf]{article}
\newcommand{\pseudo}{{P}}
\epsfverbosetrue
\newcommand{\half}{\frac{1}{2}}
\newcommand{\figurebox}[2]{\fbox{\vbox to#2in{\hbox to #1in{\hfil} \vfil}}}
\newcommand{\er}[2]{\raisebox{0.1em}{\scriptsize
			{$\;\begin{array}{@{}l@{}}
			  +\makebox[0.75em][r]{#1} \\[-0.3em]
			  -\makebox[0.75em][r]{#2}
			\end{array}$}}}
\newcommand{\err}[2]{\raisebox{0.1em}{\scriptsize
			{$\;\begin{array}{@{}l@{}}
			  +\makebox[1.25em][r]{#1} \\[-0.3em]
			  -\makebox[1.25em][r]{#2}
			\end{array}$}}}
\newcommand{\errr}[2]{\raisebox{0.1em}{\scriptsize
			{$\,\begin{array}{@{}l@{}}
			  +\makebox[1.4em][r]{#1} \\[-0.3em]
			  -\makebox[1.4em][r]{#2}
			\end{array}$}}}
\newcommand{\bm}[1]{\mbox{\boldmath ${#1}$}}
\newcommand{\beq}{\begin{equation}}
\newcommand{\eeq}{\end{equation}}
\newcommand{\dleft}{\stackrel{\leftarrow}{D}}
\newcommand{\dright}{\stackrel{\rightarrow}{D}}
\begin{document}

\begin{titlepage}

\begin{flushright}
Edinburgh Preprint: 92/507\\
Liverpool Preprint: LTH 301\\
Southampton Preprint: SHEP 92/93-13\\
hep-lat/9307009\\
Revised 30 June 1993
\end{flushright}
\vspace{5mm}
\begin{center}
{\Huge Quenched Light Hadron Mass Spectrum and Decay Constants: the
effects of $O(a)$-Improvement at $\beta=6.2$}\\[15mm] {\large\it UKQCD
Collaboration}\\[3mm]

{\bf C.R.~Allton\footnote{Present address: Dipartimento di Fisica,
Universit\`a di Roma {\em La Sapienza}, 00185 Roma, Italy},
C.T.~Sachrajda}\\
Physics Department, The University, Southampton
SO9~5NH, UK

{\bf R.M.~Baxter, S.P.~Booth, K.C.~Bowler, S.~Collins, D.S.~Henty,
R.D.~Kenway, C.~McNeile\footnote{Present address: Dept.~of
Physics \& Astronomy, University of Kentucky, Lexington KY 40506,
USA}, B.J.~Pendleton, D.G.~Richards, J.N.~Simone, A.D.~Simpson}\\
Department of Physics, The University of Edinburgh,
Edinburgh EH9~3JZ, Scotland

{\bf A.~McKerrell, C.~Michael, M.~Prisznyak,}\\
DAMTP, University of Liverpool, Liverpool L69~3BX, UK

\end{center}
\vspace{5mm}
\begin{abstract}

We compare the light hadron spectrum and decay constants for quenched
QCD at $\beta=6.2$ using an $O(a)$-improved nearest-neighbour Wilson
fermion action with those obtained using the standard Wilson fermion
action on the same set of 18 gauge configurations.  For pseudoscalar
meson masses in the range 330--800~MeV, we find no significant
difference between the results for the two actions.  The scales
obtained from the string tension and mesonic sector are consistent,
but differ from that derived from baryon masses.  The ratio of the
pseudoscalar decay constant to the vector meson mass increases
slowly with quark mass as observed experimentally.

\end{abstract}

\end{titlepage}

\section{Introduction}\label{introduction}

The numerical simulation of QCD involves systematic errors arising
from a variety of sources: non-zero lattice spacing, finite
volume, quenching and/or unphysical quark masses.  Here we address
the first of these by working in the quenched approximation at a
fixed volume, and comparing the results obtained on the same set
of gauge configurations using two formulations of the fermion
action with different discretisation errors.

Classically, the standard Wilson pure gauge action differs from
the continuum Yang-Mills action by terms of $O(a^2)$, where $a$ is
the lattice spacing~\cite{luscher-weisz}, whereas the Wilson
formulation of lattice fermions introduces an $O(a)$
discretisation error in order to avoid the doubling problem.  At
the quantum level, matrix elements computed with the Wilson action
have errors of $O(a)$.  Monte-Carlo determinations of the
renormalisation constants of the vector current have shown that
these discretisation errors may be as large as 30\%~\cite{hmprs}.

In the spirit of the Symanzik improvement programme~\cite{symanzik},
which sought to eliminate in a systematic way discretisation errors at
the quantum level, Wetzel~\cite{wetzel} proposed a two-link fermion
action which cancels the $O(a)$ term in the Wilson action at tree
level.  In a study of on-shell improvement, Sheikholeslami and
Wohlert~\cite{sheikholeslami} introduced a nearest-neighbour
$O(a)$-improved fermionic action, which is more convenient for large
scale simulations and which we refer to as the `SW' action.  This
may be obtained from the two-link action by a rotation of fermion
fields in the functional integral.  Heatlie et al.~\cite{hmprs}
demonstrated the absence of the leading logarithmic terms at $n^{\rm
th}$ order in perturbation theory, in matrix elements of rotated
operators.  These terms are of the form $g_0^{2n}a\ln^n a$, and hence
effectively $O(a)$ in the weak coupling regime, because $g_0^2\sim
1/\ln a$.

\begin{figure}[htbp]
\begin{center}
\leavevmode
\epsfxsize=200pt
\epsffile{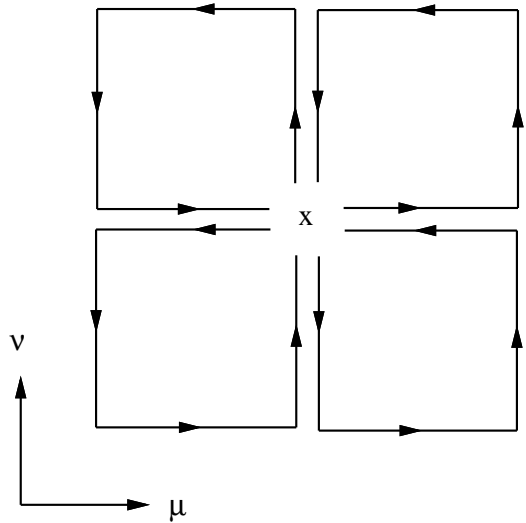}
\caption{Lattice definition of the field strength tensor,
$F_{\mu\nu}(x)$.\label{F}}
\end{center}
\end{figure}

In this paper we present further details of a study in quenched
QCD on a $24^3\times 48$ lattice at $\beta=6.2$, using both the
standard $r=1$ Wilson fermion action:
\begin{eqnarray}
S_F^W & = & \sum _x \Biggl\{ \bar{q}(x)q(x)-\kappa\sum _\mu\Biggl[
            \bar{q}(x)(1-\gamma _\mu)U_\mu (x) q(x+\hat\mu ) +
            \bar{q}(x+\hat\mu )(1+\gamma _\mu)U^\dagger _\mu(x)
            q(x)\Biggr]\Biggr\}\;\;\;\;\;
\end{eqnarray}
and the SW fermion action~\cite{sheikholeslami}:
\begin{equation}
S_F^{SW}  = S_F^W - i\frac{\kappa}{2}\sum_{x,\mu,\nu}\bar{q}(x)
         F_{\mu\nu}(x)\sigma_{\mu\nu}q(x).
\label{eq:action}
\end{equation}
$F_{\mu\nu}$ is a lattice definition of the field strength tensor,
which we take to be the sum of the four untraced plaquettes in the
$\mu\nu$ plane open at the point $x$~\cite{sheikholeslami}, as
indicated in Figure~\ref{F}:
\begin{equation}
F_{\mu\nu}(x) = \frac{1}{4} \sum_{\Box=1}^{4} \frac{1}{2i}
\biggl[U_{\Box\mu\nu}(x) - U_{\Box\mu\nu}^\dagger(x)\biggr].
\end{equation}
For this reason, the SW action is sometimes called the `clover'
action~\cite{hadrons_lett}.

Our objective is to look for evidence of improvement in masses and
decay constants of the light hadrons.  A summary of our results has
appeared in~\cite{hadrons_lett}, where we concluded that the SW
action for medium to light quark masses gives results in agreement
with the Wilson action, although typically with a noisier signal.  In
this paper, we report our attempts to enhance the signal when using the
SW action by smearing the same set of propagators at the sink.

The plan of the paper is as follows.  We begin by describing the
generation of the gauge fields, and the measurement of the static
quark-antiquark potential used to extract the lattice spacing in
physical units from the string tension.  We describe the
algorithms used to compute the quark propagators, and the
bootstrap method used to analyse the hadron correlators.  We
compare the meson and baryon spectra for the two actions, and
extract values for the lattice spacing by comparing the masses to
their physical values.  We investigate the hyperfine mass splittings
as quantities that might be sensitive to the difference between
the actions.  Finally, we compute the pseudoscalar and vector
meson decay constants.

\section{Computational Procedure}

The gauge field configurations and quark propagators were obtained
using the Meiko i860 Computing Surface at Edinburgh.  This is a MIMD
system of 64 nodes, each of which consists of one Intel i860 processor
on which the application runs, two Inmos T800's for inter-node
communications and 16~MBytes of memory.  The peak performance is
5~Gflops in 32-bit arithmetic.  Our codes are written in ANSI C with
key numerically-intensive routines in i860 assembler, and
communications using Meiko CS-Tools~\cite{kcb1,kcb2}.  Apart from
certain global operations which use 64 bits, the calculations are
performed in 32-bit arithmetic.  We achieve a maximum performance of
3~Gflops for some routines, while overall our programs sustain around
1.5~Gflops~\cite{chep92}.

\subsection{Gauge Configurations}

\subsubsection{Update algorithm}

There are computational advantages in using an over-relaxation (OR)
algorithm for updating, both from the speed of the code itself and
from the improved transport through the space of equilibrium
configurations.  For $SU(2)$ pure gauge theory, such an algorithm is
very simple and hence computationally fast.  For $SU(3)$ pure gauge
theory, there is no such simple implementation known.  The proposal of
Creutz~\cite{Cr} uses an approximate over-relaxation which has to be
corrected by an accept/reject calculation. Furthermore, a projection of
a $ 3 \times 3 $ complex matrix to $SU(3)$ is needed, which is
computationally slow.

As an alternative, we explored the possibility of using OR steps in a
sequence of $SU(2)$ subgroups of $SU(3)$.  This is in the spirit of
the Cabibbo-Marinari~\cite{C_M} scheme of updating $SU(3)$ via $SU(2)$
subgroups. Our $SU(2)$ subgroup OR scheme is exactly micro-canonical,
unlike the Creutz scheme. In order to optimise this $SU(2)$ subgroup
OR approach, we measured the average distance moved by a link in the
$SU(3)$ group manifold (along a geodesic) for different subgroup
updating schemes. Using 3 subgroups (12, 23 then 31) was optimal and
gave a distance moved slightly less than the Creutz scheme, but
significantly more than the Cabibbo-Marinari heat-bath update. A check
of auto-correlation between configurations was made and confirmed that
this $SU(2)$ OR scheme is competitive.  Because of its computational
speed, we chose to use this 3 subgroup $SU(2)$ OR algorithm,
complemented by one heat-bath update every 5 OR updates to preserve
ergodicity. We call this type of algorithm `Hybrid Over-Relaxed'.  At
the end of each $5+1$ sweeps, we re-unitarise each element of the
gauge configuration by normalising the first row of the $SU(3)$
matrix, making the second row orthonormal to the first and then
reconstructing the third row from the appropriate vector product. We
have checked that any systematic effect arising from this procedure is
negligible.

Whenever the gauge configuration is written to disk, the last
operation performed on it is a re-unitarisation.  This allows us to
write only the first two rows of each matrix and yet re-start from the
identical configuration by reconstructing the final row.

Random numbers were generated using the {\it uni\/} random number
generator~\cite{uni}. Each processor used a separate copy of the
generator, initialised with different starting states.

The results presented here are based on an analysis for each action
of the same set of 18 configurations, starting at configuration 16800
and separated by 2400 sweeps.

\subsubsection{Static quark--antiquark potential}

The potential between static colour sources at separation $R$ can be
obtained by measuring Wilson loops of size $R\times T$.  However, it
is useful to consider generalised correlations of size $R\times T$
where the spatial paths are any gauge invariant path with cylindrical
symmetry $(A_{1g})$, not just the straight line used in a rectangular
Wilson loop.  Moreover, the two spatial paths, at times $T$ apart,
may be different.  So the measurable is $C_{ij}(R,T)$ where $i,j$
label the path type.  Then the largest eigenvalue, $\lambda (R)$, of the
transfer matrix in the presence of the static sources is related to
the potential by
\beq
V(R) = - \log ( \lambda (R) )
\eeq
and $\lambda (R)$ can be obtained from the limit as $T\to\infty $ of
the largest eigenvalue, $\lambda (R,T)$, of the equation
\beq
C_{ij}(R,T) u_{j} = \lambda (R,T) C_{ij}(R,T-1) u_{j}.
\eeq
This is a variational method where an optimal combination of paths $i$
(with $i=1,..N)$ is chosen.  The best choice of paths would allow
$\lambda (R,T)$ to be close to its asymptotic value for small $T$.  In
this way the statistical errors are minimised.  Thus it is convenient
to introduce an `overlap' defined as
\beq
{\cal O}(R) = \prod_{T}\{ \lambda (R,T)/\lambda (R) \}.
\eeq
For $R=12$ straight paths, we find ${\cal O}<0.006.$ Thus rectangular
Wilson loops are a very inaccurate way of extracting the ground state
potential.

As has been known for some time, an efficient improvement comes from
fuzzing or blocking these paths~\cite {T,APE,PM}.  Using a purely
spatial blocking allows the transfer matrix interpretation to be
retained.  Because of its flexibility, we used the recursive blocking
scheme~\cite {APE}:
\beq
U(\hbox{new}) = {\cal P}_{SU(3)}\Bigl[ c
U(\hbox{straight}) + \sum^{4}_{1} U(\hbox{u-bends})\Bigr].
\eeq
After exploratory studies to optimise the overlap ${\cal O}$, we chose
$c=4$ and a maximum of $40$ iterations of this recursive blocking.  As
a variational basis with $N=2,$ we used $40$ and $28$ iterations for
the paths.  This basis for $R=12$ yields an overlap ${\cal O}=0.95$
which is a huge improvement over the purely unblocked case.  Because
this value of the overlap is close to $1.0$, one can get reliable
estimates of the ground state potential from quite small $T$-values
such as $T=3$. A powerful cross check of this extraction of the
ground state energy at rather small $T$-values comes from determining
the first excited state also in the variational method.  From previous
work, this first excited $A_{1g}$ potential should lie about $2\pi /R$
higher~\cite{PM} and this is completely consistent with our results.

To estimate the potential $V(R)$, we need to extrapolate in $T$.  For
$1 \le R \le 4$, we find that the $3:2$ and $4:3$ $T$-ratio results
for $\lambda (R,T)$ are consistent within errors. As a conservative
estimate of the asympotic value we used the $4:3$ $T$-ratio to evaluate
$V(R)$.  For $R \ge 5$, we find a small but statistically significant
difference between the $3:2$ and $4:3$ $T$-ratio results. This we
extrapolate to large $T$ using the estimate of the first excited state
energy referred to above.  Statistical errors come from a bootstrap
analysis of the variation over our $18$ sample configurations, where
the variational path combination is fixed by the $1:0$ $T$-ratio
analysis.  This avoids the possibility that the variational approach
is influenced by the statistical error.  Results are shown in
Table~\ref{potential}.
\begin{table}
\begin{center}
\begin{tabular}{|c|l|c|}\hline
 $R$ \  & $\ V(R) $  & $V(R+1)-V(R)$   \\\hline
 1    & 0.3775(2)      &0.1564(3)      \\
 2    & 0.5339(3)      &0.0787(5)      \\
 3    & 0.6126(7)      &0.0509(5)      \\
 4    & 0.6635(8)      &0.0416(12)     \\
 5    & 0.7043(17)     &0.0348(14)     \\
 6    & 0.7392(26)     &0.0316(19)     \\
 7    & 0.7707(27)     &0.0300(19)     \\
 8    & 0.8008(39)     &0.0304(24)     \\
 9    & 0.8311(37)     &0.0295(23)     \\
 10   & 0.8606(53)     &0.0212(29)     \\
 11   & 0.8819(60)     &0.0322(26)     \\
 12   & 0.9140(60)   &                   \\\hline
\end{tabular}
\caption{ The potential and force at $\beta =6.2$.\label{potential}}
\end{center}
\end{table}

We  also  fit  the  static  potential  for $R\ge 2$   with   a lattice
Coulomb plus linear term:
\beq
V(R) = C - {E\over R_{L}}  + K R
\eeq
where $1/R_{L}$ is the discrete Coulomb Green function.  Using a fit
to the previously determined values of $V(R)$ and taking into account
correlations in errors between potentials at different $R$-values, we
find the result of Table~\ref{force}.  Here the error quoted is
statistical; it includes the error from the data sample variation and
the $T$-extrapolation described above, but not the systematic effect of
using different fit functions. This value for the string tension
enables us to set the scale by requiring $a^{-1}\sqrt K=0.44$ GeV, so
yielding $a^{-1} = 2.73(5)$ GeV. This implies that the potential has
been measured to a physical distance, $Ra$, of 0.87 fermi.  Using the
conventional 2-loop perturbative relationship between $\Lambda$ and
$a$ yields $\sqrt K/a\Lambda _{L} \le 85.9(1.5)$, where the inequality
arises because of the observed lack of asymptotic scaling at
$\beta=6.2$.
\begin{table}
\begin{center}
\begin{tabular}{|c|c|c|}\hline
 $E$ & $K$  \\\hline
  0.274(6) & 0.0259(9) \\\hline
\end{tabular}
\caption{Fit to the force for $R>1$.\label{force}}
\end{center}
\end{table}

These results are in agreement with a similar analysis on $20^{4}$
lattices at the same $\beta $-value~\cite {PM,MT}.  This confirms the
conclusion of that work that finite-size effects are small for the
potential on lattices with $L~\ge ~20$. A recent independent analysis
on a $24^{3} \times 32$ lattice~\cite{BS} gives a result for the
string tension which also agrees with ours.  From a comparison of our
result for the string tension with those at neighbouring $\beta$
values, we find that asymptotic scaling is not obeyed.  This can be
understood because the lattice bare coupling is a poor expansion
parameter~\cite{lm}. The study of large lattices at larger $\beta$
confirms this~\cite{ukqcd_su3} and shows that the use of a more
physical coupling gives excellent agreement with perturbation theory
to two loops, allowing $\Lambda$ to be related to the string tension.

\subsection{Quark Propagators}

To calculate the quark propagator we need to solve equations of
the form
\beq ({\cal A} - \kappa {\cal B}) G(x,0) = \eta_{x,0}.
\label{eq:Mpsieqeta}
\eeq
where, for either action, ${\cal A}$ is purely local and ${\cal
B}$ connects only nearest-neighbour sites.  We have investigated
the least residual (LR) and least norm (LN)
variants~\cite{oyanagi} of the Conjugate Gradient~(CG) algorithm,
and an Over-Relaxed Minimal Residual~(MR)
algorithm~\cite{hockney,simpson}.  We also implemented a simple
preconditioning which involves left-multiplying
Equation~(\ref{eq:Mpsieqeta}) by \mbox{$(1+\kappa{\cal
BA}^{-1})$}.  This decouples even and odd lattice sites so that
Equation~(\ref{eq:Mpsieqeta}) becomes
\begin{eqnarray}
\label{eq:even_prop}
({\cal A} - \kappa^2 {\cal BA}^{-1}{\cal B}) G^{\rm even}
& = & \eta^{\rm even} + \kappa {\cal BA}^{-1}\eta^{\rm odd}\\
\label{eq:odd_prop}
G^{\rm odd} & = & {\cal A}^{-1} \eta^{\rm odd}
                  + \kappa {\cal A}^{-1}{\cal B} G^{\rm even}.
\end{eqnarray}
We solve Equation~(\ref{eq:even_prop}) for $G^{\rm even}$ and
substitute into Equation~(\ref{eq:odd_prop}) to obtain $G^{\rm
odd}$.  For all the algorithms and for both actions, we found that
this red-black preconditioning gave a factor of between 2.5 and
3.0 decrease in the number of iterations required, without
increasing significantly the number of floating-point operations
per iteration.

Writing
\begin{eqnarray}
{\cal M}  & = & {\cal A} - \kappa^2 {\cal BA}^{-1}{\cal B}\\
\zeta     & = & \eta^{\rm even} + \kappa {\cal BA}^{-1}\eta^{\rm
odd}
\end{eqnarray}
the MR algorithm is
\begin{eqnarray}
r_0                & = & \zeta - {\cal M} G^{\rm even}_0
\nonumber \\
\mbox{repeat until convergence}
\nonumber \\
\alpha             & = & \frac{({\cal M} r_i, r_i)}{({\cal M} r_i,
{\cal M} r_i)}
\nonumber \\
\alpha'            & = & \Omega \alpha \hspace{10mm}
\mbox{(over-relaxation)}
\nonumber \\
G^{\rm even}_{i+1} & = & G^{\rm even}_i + \alpha' r_i
\nonumber \\
r_{i+1}            & = & r_i - \alpha' {\cal M} r_i.
\nonumber
\end{eqnarray}
We found that choosing the over-relaxation parameter, $\Omega$, to
be 1.1 gave a gain from over-relaxation of a few per cent.  This
is in agreement with the data in \cite{hockney}, which indicates
that the gains from over-relaxation decrease as $\beta$ increases.
However, for both actions and throughout the physical regime
investigated, MR typically converges to a solution in fewer
iterations than CG, with each iteration taking about half the
time.  Therefore, we always use an over-relaxed MR algorithm with
red-black preconditioning.

In order to obtain an improved matrix element of an operator
containing quark fields, in addition to using the SW action,
the quark fields must be rotated according to~\cite{hmprs,MSV}:
\begin{eqnarray}
q & \rightarrow & q^{\prime} = \left(
1-\frac{1}{2}\gamma\cdot\dright\right)\,q\label{eq:psiimp}\\
\bar q & \rightarrow & \bar q^{\prime} = \bar q\,
\left( 1+\frac{1}{2}\gamma\cdot\dleft\right)
\label{eq:psibimp}
\end{eqnarray}
where we have used the equation of motion for the quark fields to
remove the bare mass from the rotation.  The lattice covariant
derivatives are defined as follows:
\beq
\dright_\mu f(x) = \half \left( U_\mu(x) f(x+\hat\mu ) -
U^\dagger_\mu(x-\hat\mu) f(x-\hat\mu) \right)
\label{eq:dra}
\eeq
and
\beq
f(x) \dleft_\mu = \half \left( f(x+\hat\mu) U^\dagger_\mu(x) -
f(x-\hat\mu) U_\mu (x-\hat\mu) \right).
\label{eq:dla}
\eeq
We can define the rotated SW propagator, $G^R(x,y)$,
by~\cite{MSSV}
\beq
G^R(x,y) \equiv (1-\frac{1}{2}\gamma\cdot\dright) G(x,y)
(1+\frac{1}{2}\gamma\cdot\dleft).
\label{eq:simp}
\eeq
This propagator includes the rotations (\ref{eq:psiimp}) and
(\ref{eq:psibimp}) and so the evaluation of all correlation
functions is performed in the same way as with the unimproved
action; no further rotations of the operators are necessary.  The
evaluation of the rotated propagator, $G^R$, is only marginally more
complicated than that of $G$; solve the equation:
\beq
({\cal A} - \kappa {\cal B}) G'(x,y) = \eta _{x,y}
+ \frac{1}{2}\eta_{x,y}\gamma\cdot\dleft
\label{eq:s1inv}
\eeq
then $G^R$ is obtained easily from $G'$ by
\beq
\label{eq:GR}
G^R(x,y)=(1-\frac{1}{2}\gamma\cdot\dright) G'(x,y).
\eeq

Recently, there has been increasing evidence that lattice measurements
can be enhanced through the use of non-local, or smeared
operators~\cite{MT,Wuppertal,LANL}.  The original attempts to use
smeared operators for fermions are described in
references~\cite{kenway84,billoire85}. For propagator calculations, we
can use a source, $\eta$, in Equation~(\ref{eq:Mpsieqeta}), which is
extended over some non-zero spatial volume, called source smearing, or
we can smear the solution, called sink smearing.  G\"{u}sken et
al.~\cite{Wuppertal} have proposed a gauge-invariant smearing method
based upon using the solution of the three-dimensional Klein-Gordon
equation: %
\beq
(1 - \kappa_S \bm{D}^2)\eta_{x,0} = \delta_{x,0}
\eeq
for the source in Equation~(\ref{eq:Mpsieqeta}), where $\bm{D}^2$ is
the three-dimensional operator
\begin{equation}
\bm{D}^2 f(x) = \sum_{j=1}^3 \left( U_j(x) f(x+\hat{j}) +
U^\dagger_j(x-\hat{j}) f(x-\hat{j})\right).
\end{equation}
Alternatively, sink smearing corresponds to solving
\beq
(1 - \kappa_S \bm{D}^2) G_{SL} = G^R
\eeq
where $G^R$ is the solution of Equation~(\ref{eq:GR}).  The extent
of the smearing is controlled by the single parameter, $\kappa_S$.

We label propagators with a local source and sink, LL, those with a
local source but smeared sink, SL, and those with a smeared source but
local sink, LS.  In practice, we smear at the sink, as this has the
advantage that the smearing can be undone with a single multiplication
by $(1 - \kappa_S \bm{D}^2)$.  Since we can thus easily recover the LL
propagator from the SL propagator, we only write out the SL propagator
to disk, a very significant advantage for our particular machine
architecture.  Details of our investigations of the effects of
smearing are reported elsewhere~\cite{smearing}.

We have calculated LL propagators at $\kappa = $ 0.1510, 0.1520,
0.1523, 0.1526 and 0.1529 for the Wilson action, LL and SL propagators
at $\kappa = $ 0.14144, 0.14226, 0.14244, 0.14262 and 0.14280 for the
SW action; the latter values were chosen to match roughly the
pseudoscalar meson masses computed in the Wilson case.  The boundary
conditions used were periodic in space and antiperiodic in time.  The
scalar hopping parameter, $\kappa_S$, used in the computation of the
SL SW propagators, was taken to be 0.180, corresponding to a smearing
radius~\cite{smearing} of approximately 2. The extra computational
cost involved in using the SW action is difficult to quantify, being
machine dependent. For the red-black preconditioned MR algorithm which
we employed, each iteration took roughly 35\% longer than for the
Wilson action, but taking into account other fixed overheads, most
notably I/O, the total elapsed time for a propagator calculation at a
given pion mass was between 10\% and 20\% longer, whilst the memory
requirement was some 10\% more than in the Wilson case.

\subsection{Analysis of 2-Point Functions}

We obtain the amplitudes and masses of mesons (baryons) by
correlated least-$\chi^2$ fits of a single cosh (exponential)
function simultaneously to the appropriate forward and backward
propagators.  We require that the fitting region, $[t_{\rm min},
t_{\rm max}]$, satisfies the condition that, within the small
range allowed by our statistics, changing $t_{\rm min}$ gives the
same mass within errors, while $t_{\rm max}$ is taken as large as
the growing noise/signal ratio allows.  We find that the intervals
$[12, 16]$ for local sinks and $[9, 13]$ for smeared sinks are
satisfactory in this regard.  The correlated $\chi^2 / \mbox{{\rm
dof}}$ varies between 0.3 and 4, which indicates both that we are
taking correlations into account and that we are getting
reasonable fits.

The covariance matrix is estimated using a single-removal
jackknife.  For simple averages, such as the propagator, this is
the same as the data covariance matrix.  We calculate the
jackknife covariance matrix for the timeslices used in the fit.
We employ a singular value decomposition in the construction of
our matrix inverses, to detect singular or nearly singular
matrices.  Because of our limited statistics, we use timeslices
12, 14, 16, 32, 34 and 36 for local sinks, and 9, 11, 13, 35, 37
and 39 for smeared sinks.  We currently have too few
configurations to deal with the $\kappa$ -- $\kappa$ correlations
directly in the fits.  Therefore we do individual fits for each
$\kappa$ value.  We recover the effect of the $\kappa$
correlations in our analysis of the errors.

To estimate our errors, we use the bootstrap resampling method, a
concise description of which has been given by Chu et
al.~\cite{bootstrap}. From our set of 18 Monte Carlo samples of the
probability distribution for the link variables, we construct a
bootstrap sample by drawing 18 configurations independently and with
replacement.  We perform exactly the same analysis on this sample of
18 configurations as on the original set, to obtain a bootstrap
estimate of average quantities.  We then build up the bootstrap
distribution by drawing 1000 bootstrap samples with corresponding
bootstrap estimates of averages.  We reconstruct the covariance matrix
for each bootstrap sample so as to take into account the uncertainty
in the covariance matrix. We then bin the bootstrap determinations of
the average to determine confidence limits.  The quoted errors are
obtained by requiring that the central 68\% of the bootstrap values
lie within the error bars.

For every quantity fitted, we use the same sequence of bootstrap
samples and store the best fit to the original set, together with the
bootstrap determinations. Thus the $i$'th bootstrap estimate of any
quantity, is the result of a fit to the $i$'th sample of 18
configurations.  In this way we preserve information on correlations
between physical quantities at, say, different $\kappa$ values.

\section{Hadron Spectrum}

\subsection{Masses and Matrix Elements}

We present results for the pseudoscalar meson (P), vector meson (V),
nucleon (N) and $\Delta$ using the following local interpolating
fields:
\begin{eqnarray}
P   &=& \bar{u}\gamma_5 d\\
V_i &=& \bar{u}\gamma_i d\\
N   &=& \epsilon_{abc}(u^aC\gamma_5 d^b)u^c\\
\Delta_\mu &=& \epsilon_{abc}(u^aC\gamma_\mu u^b)u^c.
\end{eqnarray}
For the vector meson, we average our correlators over the three
polarisation states, for the nucleon we average the 11 and 22 spinor
indices of the correlator, and for the $\Delta$ we project out the
spin-$3/2$ component and average over the four spin projections.

The mass estimates in lattice units obtained for the Wilson action
using local sources and sinks are given in
Table~\ref{tab:WLL_masses}.  Also included are the amplitudes, $A$,
and $\chi^2$/dof obtained in the single cosh or exponential fits
to the zero-momentum timeslice propagators:
\begin{equation}
\sum_{\bm{x}}\langle h(\bm{x},t)h^\dagger(0)\rangle
\sim \left\{ \begin{array}{ll}
        A_h\; (e^{-m_ht} + e^{-m_h(L_t-t)}),   & {\rm (mesons)}\\
        A_h\; e^{-m_ht},\hspace{5mm}t < L_t/2, & {\rm (baryons)}
        \end{array}\right.
\end{equation}
described above, where $L_t=48$ is the total time extent of our
lattice.  The corresponding results for the SW action using
local sources with both local and smeared sinks are
given in Table~\ref{tab:CLL_CSL_masses}.
\begin{table}
\begin{center}
\begin{tabular}{|c|l|l|l|l|l|l|}
\hline
\multicolumn{7}{|c|}{Wilson (LL)}\\\hline
$\kappa$
& \multicolumn{1}{|c|}{0.1510}
& \multicolumn{1}{|c|}{0.1520}
& \multicolumn{1}{|c|}{0.1523}
& \multicolumn{1}{|c|}{0.1526}
& \multicolumn{1}{|c|}{0.1529}
& \multicolumn{1}{|c|}{0.15328\er{7}{4}}
\\
\hline
$m_{P}$ & 0.295\err{6}{3} & 0.221\err{9}{3} & 0.195\err{9}{3} &
0.164\err{11}{4}
& 0.122\err{9}{11} & 0.0 \\ $A_{P} \times 10^1$ & 0.121\err{8}{7} &
0.118\err{9}{7} & 0.124\err{10}{7} & 0.140\err{15}{10} &
0.180\err{25}{46} & \\
$\chi^2$/dof & 2.1/4 & 1.8/4 & 1.5/4 & 1.8/4 & 2.0/4 & \\\hline $m_V$ &
0.377\err{11}{5} & 0.332\err{15}{4} & 0.321\err{16}{6} &
0.310\err{19}{11} &
0.298\err{39}{34} & 0.277\err{25}{9}\\ $A_V \times 10^1$ &
0.130\err{19}{13} &
0.102\err{19}{7} & 0.095\err{2}{1} & 0.089\err{2}{1} &
0.086\err{4}{2}& \\
$\chi^2$/dof & 4.5/4 & 6.6/4 & 7.4/4 & 7.3/4 & 10.0/4 & \\\hline
$m_N$ &
0.591\err{11}{6} & 0.509\err{15}{7} & 0.480\err{16}{9} &
0.445\err{16}{13} &
0.395\err{48}{42} & 0.393\err{20}{16}\\ $A_N \times 10^4$ &
0.217\err{33}{22} &
0.132\err{32}{16} & 0.109\err{27}{15} & 0.085\err{24}{15} &
0.062\err{47}{23} &
\\ $\chi^2$/dof & 1.2/4 & 5.1/4 & 5.5/4 & 6.4/4 & 12.4/4 & \\\hline
$m_\Delta$ &
0.647\err{19}{10} & 0.582\err{23}{9} & 0.560\err{23}{11} &
0.538\err{36}{18} &
0.533\errr{116}{40} & 0.496\err{31}{16}\\ $A_\Delta \times 10^3$ &
0.109\err{27}{15} & 0.065\err{19}{9} & 0.051\err{14}{7} &
0.039\err{20}{8} &
0.039\errr{104}{14} & \\ $\chi^2$/dof & 1.4/4 & 2.5/4 & 6.2/4 &
8.4/4 & 5.6/4 &
\\\hline
\end{tabular}
%
%
\caption{Hadron masses and amplitudes, in lattice units, for the
Wilson action.  The last column contains the values obtained by
linear extrapolation to $m_{P}=0$.\label{tab:WLL_masses}}
\end{center}
\end{table}
\begin{table}
\begin{center}
\begin{tabular}{|c|l|l|l|l|l|l|}
\hline
\multicolumn{7}{|c|}{SW (LL)}\\\hline
$\kappa$
& \multicolumn{1}{|c|}{0.14144}
& \multicolumn{1}{|c|}{0.14226}
& \multicolumn{1}{|c|}{0.14244}
& \multicolumn{1}{|c|}{0.14262}
& \multicolumn{1}{|c|}{0.14280}
& \multicolumn{1}{|c|}{0.14313\er{7}{4}}
\\
\hline
$m_{P}$    & 0.302\err{6}{4}   & 0.217\err{8}{6}  & 0.194\err{9}{6}
& 0.168\err{10}{6}     & 0.135\err{11}{6}  & 0.0 \\
$A_{P} \times 10^1$    & 0.145\err{11}{12} & 0.139\err{12}{14}
& 0.145\err{13}{15}    & 0.157\err{19}{17} & 0.190\err{31}{25} & \\
$\chi^2$/dof & 3.0/4   & 1.8/4   & 1.4/4   & 1.4/4  & 2.5/4   & \\\hline
$m_V$      & 0.395\err{13}{9}  & 0.345\err{18}{10} & 0.338\err{23}{13}
& 0.329\err{30}{23} & 0.313\err{35}{44} & 0.292\err{26}{21} \\
$A_V \times 10^1$& 0.103\err{18}{17} & 0.079\err{19}{10}
& 0.077\err{21}{11} & 0.071\err{25}{15} & 0.060\err{24}{22} & \\
$\chi^2$/dof & 6.3/4& 7.6/4   & 6.7/4   & 5.9/4   & 6.7/4  & \\\hline
$m_N$      & 0.608\err{15}{8}  & 0.495\err{30}{8} & 0.460\err{35}{11}
& 0.419\err{45}{17}  & 0.396\err{51}{39} & 0.375\err{42}{17}\\
$A_N \times 10^4$    & 0.196\err{56}{25} & 0.091\err{45}{13}
& 0.070\err{38}{12}  & 0.053\err{37}{12} & 0.052\err{46}{21} & \\
$\chi^2$/dof& 1.7/4  & 2.2/4   & 2.2/4   & 3.1/4   & 5.7/4  & \\\hline
$m_\Delta$  & 0.678\err{15}{12} & 0.598\err{25}{15} & 0.577\err{25}{22}
& 0.565\err{33}{33}   & 0.586\err{89}{62} & 0.513\err{40}{31}\\
$A_\Delta \times 10^3$& 0.080\err{16}{15} & 0.037\err{11}{6}
& 0.029\err{9}{8}     & 0.026\err{12}{9}  & 0.036\err{64}{19} & \\
$\chi^2$/dof& 1.2/4 & 11.4/4  & 11.1/4  & 7.5/4   & 3.2/4  & \\\hline\hline
\multicolumn{7}{|c|}{SW (SL)}\\\hline
$\kappa$
& \multicolumn{1}{|c|}{0.14144}
& \multicolumn{1}{|c|}{0.14226}
& \multicolumn{1}{|c|}{0.14244}
& \multicolumn{1}{|c|}{0.14262}
& \multicolumn{1}{|c|}{0.14280}
& \multicolumn{1}{|c|}{0.14311\er{6}{3}}
\\
\hline
$m_{P}$        & 0.304\err{6}{5}  & 0.216\err{7}{5}  & 0.194\err{7}{5}
& 0.169\err{8}{6}     & 0.136\err{13}{8} & 0.0 \\
$A_{P}\times 10^{-1}$ & 0.716\err{59}{66}& 0.636\err{47}{46}
& 0.648\err{41}{45}   & 0.693\err{42}{50}& 0.836\err{97}{98} & \\
$\chi^2$/dof& 6.4/4   & 5.0/4   & 5.0/4  & 4.7/4   & 3.8/4  & \\\hline
$m_V$          & 0.398\err{9}{5}   & 0.349\err{13}{11} & 0.338\err{16}{15}
& 0.325\err{21}{19}    & 0.310\err{39}{29} & 0.299\err{20}{18} \\
$A_V\times 10^{-1}$ & 0.831\err{79}{59} & 0.650\err{77}{67}
& 0.600\err{86}{73}    & 0.542\errr{114}{83}  & 0.484\errr{216}{120} & \\
$\chi^2$/dof& 2.8/4    & 2.1/4    & 2.3/4  & 3.0/4   & 4.9/4   & \\\hline
$m_N$          & 0.609\err{13}{9}  & 0.503\err{11}{13} & 0.474\err{13}{15}
& 0.447\err{15}{22} & 0.408\err{26}{24} & 0.383\err{19}{26}\\
$A_N$               & 0.308\err{55}{58} & 0.150\err{24}{28}
& 0.127\err{25}{33} & 0.118\err{25}{41} & 0.108\err{28}{30} & \\
$\chi^2$/dof& 11.1/4& 13.7/4   & 15.5/4 & 16.6/4   & 15.8/4 & \\\hline
$m_\Delta$     & 0.664\err{12}{8}  & 0.598\err{18}{11} & 0.582\err{24}{13}
& 0.567\err{36}{17}     & 0.549\err{60}{32} & 0.528\err{24}{22}\\
$A_\Delta\times 10^{-1}$& 0.143\err{17}{17} & 0.094\err{20}{13}
& 0.085\err{24}{13}     & 0.078\err{30}{14} & 0.071\err{41}{17}  & \\
$\chi^2$/dof & 1.6/4    & 1.1/4   & 1.4/4   & 2.7/4   & 6.3/4   & \\
\hline
\end{tabular}
%
%
\caption{Hadron masses and amplitudes, in lattice units, for the SW action.
The last column contains the values obtained by linear extrapolation to
$m_{P}=0$.\label{tab:CLL_CSL_masses}}
\end{center}
\end{table}

\begin{figure}[htbp]
\begin{center}
\leavevmode
\epsfysize=300pt
  \epsfbox[20 30 620 600]{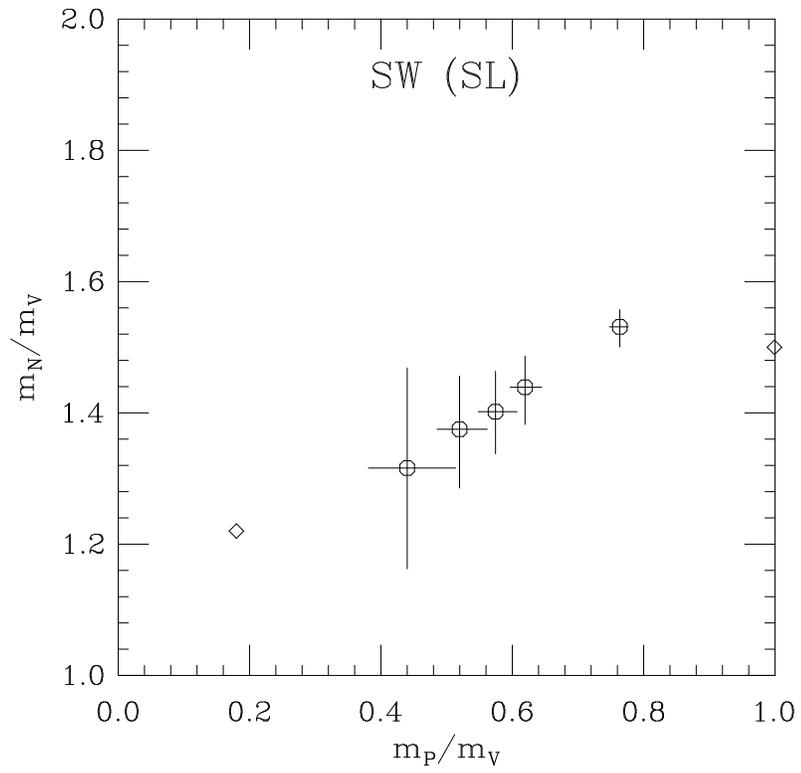}
\end{center}
\caption{Edinburgh plot for the SW action using sink smearing.}
\label{CSL_edin_plot}
\end{figure}

The Edinburgh plots for the LL data for both actions are given
in~\cite{hadrons_lett}.  That for the SW SL results is plotted in
Figure~\ref{CSL_edin_plot}.  The plots are broadly consistent, showing
a trend towards the experimental value for $m_N/m_\rho$ with
decreasing pseudoscalar meson mass. However, a comparison of the
SW LL and SL plots reveals that the errors in the SL data are
smaller than those of the LL data by at least 50\%.  This reduction is
attributable to the better signal for the nucleon correlator using the
SL data; comparing the results for the baryon masses at the lightest
quark mass for LL and SL in Table~\ref{tab:CLL_CSL_masses}, we note
that the signal increases from approximately $2\sigma$ to around
$4\sigma$ from zero.  For this reason, we employ SL data, wherever
possible, in making our comparisons of the SW and Wilson hadron
masses.

\subsection{Chiral Extrapolations}

We now consider chiral extrapolation of the hadron masses in order
to estimate the lattice scale for the two actions.

\begin{figure}[htbp]
\begin{center}
\leavevmode
\epsfysize=300pt
  \epsfbox[20 30 620 600]{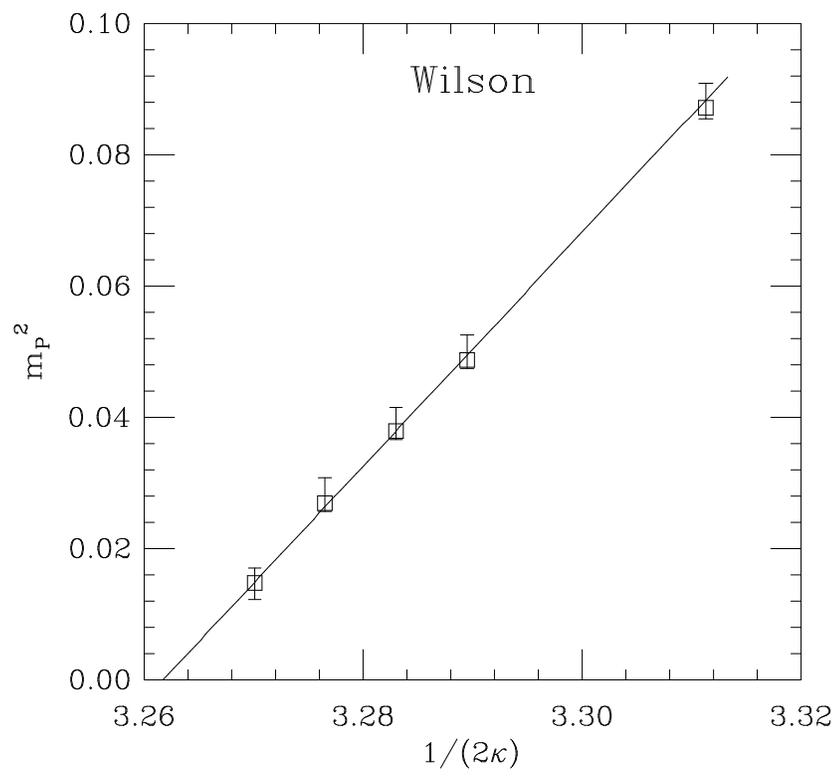}
\end{center}
\begin{center}
\leavevmode
\epsfysize=300pt
  \epsfbox[20 30 620 600]{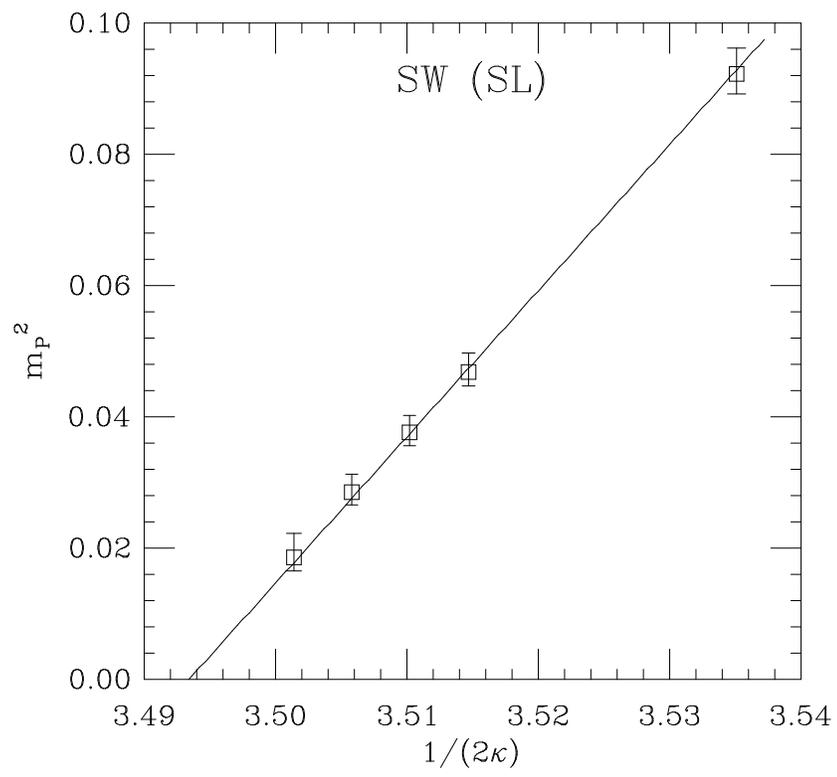}
\end{center}
\caption{$m_{P}^2$ versus $1/2\kappa$.}
\label{pcac}
\end{figure}

{}From the bootstrap analysis, we find that there are strong
correlations between the pseudoscalar meson masses at different
$\kappa$ values.  Figure~\ref{pcac} shows correlated linear fits
to the data for $m_{P}^2$ versus $1/2\kappa$ for both actions,
at all five $\kappa$ values, confirming PCAC behaviour throughout
the quark mass range used.
The strong correlations presumably account for the remarkable
linearity of our best estimates for $m_{P}^2$ as functions of
$1/2\kappa$, given the size of the statistical errors.  From the
chiral extrapolation, we obtain
\begin{eqnarray}
\kappa_{crit} &=& 0.15328\mbox{\er{7}{4}}\qquad \mbox{(Wilson)},\nonumber\\
\kappa_{crit} &=& 0.14313\mbox{\er{7}{4}}\qquad \mbox{(SW LL)},\\
\kappa_{crit} &=& 0.14311\mbox{\er{6}{3}}\qquad \mbox{(SW SL)}.\nonumber
\end{eqnarray}

The lattice scales obtained from correlated linear extrapolations
of the other hadron masses to the chiral limit are presented in
Table~\ref{tab:scales}.
\begin{table}
\centering
\begin{tabular}{|c|c|c|c|}
\hline
            & \multicolumn{3}{c|}{$a^{-1}$(GeV)}  \\
\cline{2-4}
physical quantity & \makebox[60pt][c]{Wilson} & \makebox[60pt][c]{SW (LL)}
            & \makebox[60pt][c]{SW (SL)} \\
\hline
$m_\rho$    & 2.77\err{9}{23}  & 2.63\err{21}{22} & 2.57\err{16}{16} \\
$m_N$       & 2.39\err{10}{12} & 2.50\err{12}{25} & 2.45\err{18}{11} \\
$m_\Delta$  & 2.48\err{8}{15}  & 2.40\err{16}{17} & 2.33\err{10}{10} \\
$(m_{K^\ast}-m_\rho)/(m_K^2-m_\pi^2)$
      & 2.05\err{25}{46} & 2.06\err{46}{48} & 1.93\err{34}{33} \\ \hline
\rule[-0.5em]{0em}{1.8em} $\sqrt{K}$ &
\multicolumn{3}{c|}{2.73\err{5}{5}} \\ \hline
\end{tabular}
\caption{Scales determined from different physical quantities.}
\label{tab:scales}
\end{table}
 %
 %
For both actions, the scales derived from $m_\rho$ agree well with
the scale from the string tension, whereas the scales from the
baryon masses, whilst consistent with one another, are lower, by
$2-3\sigma$ for the SW data and more than $3\sigma$ for the
Wilson data.  We have investigated quadratic chiral
extrapolations, but find that we can attach no statistical
significance to any difference between the extrapolations.

The value of $\kappa$ corresponding to the strange quark may be
estimated by assuming that the pseudoscalar meson mass obeys
\begin{equation}
m_{P}^2(\kappa_1,\kappa_2) = b_{P}\left( \frac{1}{2\kappa_1}+\frac{1}
{2\kappa_2}                            - \frac{1}{\kappa_{crit}}\right)
\label{quark_mass}
\end{equation}
for valence quark masses corresponding to $\kappa_1$ and $\kappa_2$.
We determine $b_{P}$ and $\kappa_{crit}$ from the data for degenerate
valence quarks, and match $m_{P}(\kappa_s,\kappa_{crit})$ to the
physical kaon mass, taking the scale from $m_\rho$. This gives
\begin{equation}
\kappa_s = 0.1517\mbox{\er{1}{3}} \hspace{0.3cm} \mbox{(Wilson)},\hspace{1cm}
\kappa_s = 0.1418\mbox{\er{2}{2}} \hspace{0.3cm} \mbox{(SW SL)}.
\label{eq:kappa_s}
\end{equation}

\subsection{Mass Splittings}

In an attempt to highlight any differences arising from use of the SW
action, we present estimates for the vector-pseudoscalar meson and
$\Delta$-nucleon mass splittings, which in QCD-inspired quark models
arise from spin interactions and therefore may be corrected at $O(a)$ by
the spin term in the SW action.  The suggestion that the
vector-pseudoscalar mass splitting might be susceptible to lattice
artefacts has been made previously by the APE collaboration in a
comparative study~\cite{APE_hyperfine} of Wilson and staggered fermion
actions, albeit at stronger coupling than the present work.

\begin{figure}[htbp]
\begin{center}
\leavevmode
\epsfysize=300pt
  \epsfbox[20 30 620 600]{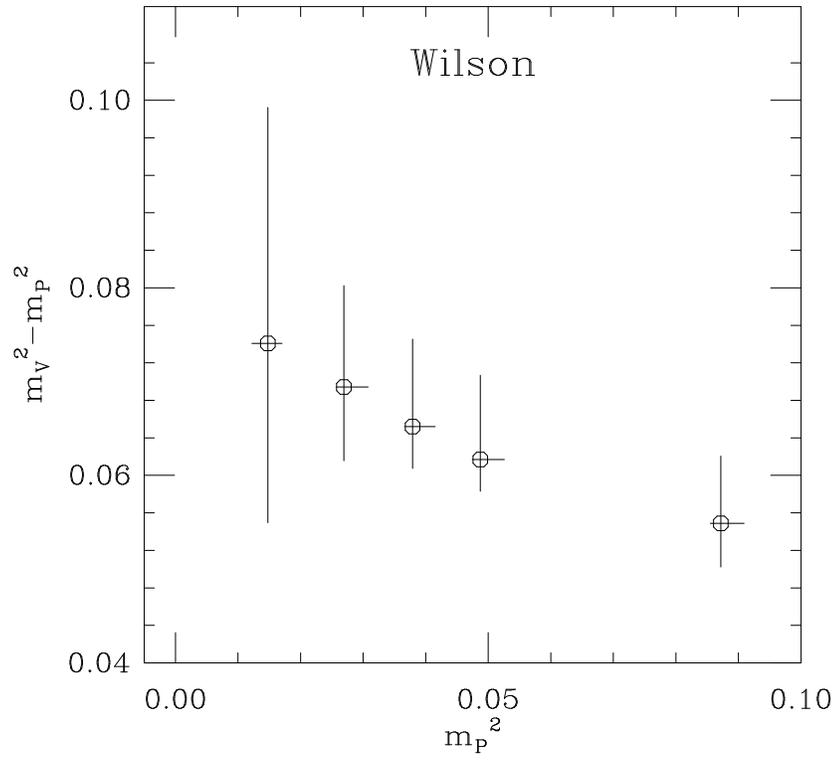}
\end{center}
\begin{center}
\leavevmode
\epsfysize=300pt
  \epsfbox[20 30 620 600]{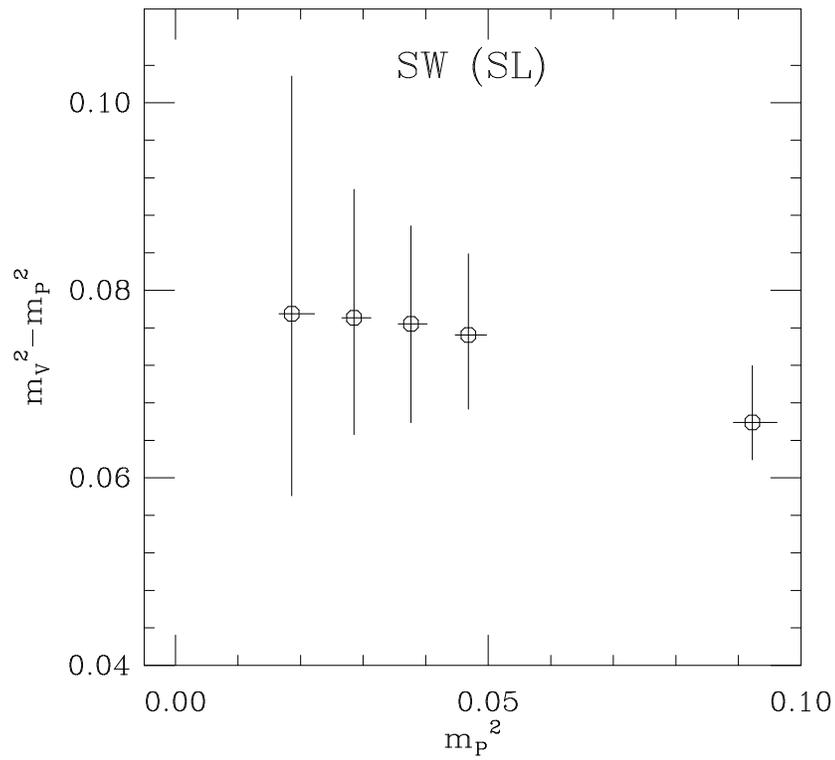}
\end{center}
\caption{$m_V^2-m_{P}^2$ versus $m_{P}^2$, in lattice units.}
\label{vec-pseudo}
\end{figure}

Experimentally, for both light-light and heavy-light mesons,
$m_V^2-m_{P}^2$ is very nearly independent of quark mass:
0.57~GeV$^2$ ($\rho-\pi$), 0.55~GeV$^2$ ($K^\ast-K$), 0.55~GeV$^2$
($D^\ast-D$), whereas for the charmonium system
$m_{J/\psi}^2-m_{\eta_c}^2 = 0.72$~GeV$^2$, suggesting that for
systems composed of equal-mass quarks, this quantity increases slowly
with quark mass.  Taking the scale from the string tension, the range
of pseudoscalar meson masses for which we have data is 330~MeV to
800~MeV.  Thus, it is of interest to study the quark-mass dependence
of our data for $m_V^2-m_{P}^2$, with equal-mass quarks.  In
Figure~\ref{vec-pseudo}, we plot the quantity $m_V^2-m_{P}^2$,
calculated directly from the bootstrap masses, versus $m_{P}^2$, for
both actions.  The Wilson data is consistent with previous
work~\cite{hyperfine} which indicated a negative slope, inconsistent
with experiment at large quark mass.  The larger statistical
errors in the SW data leave open the possibility of a reduced
dependence on quark mass, although it appears that there is still
a tendency for the hyperfine splitting to decrease with increasing
quark mass.  We have shown in a related
study~\cite{UKQCD_hyperfine} that at heavier quark masses this is
indeed the case; the SW estimate of the hyperfine splitting in
charmonium is a factor of 1.8 larger than the Wilson estimate,
although still smaller by a factor of roughly two than the
experimental value.  Using the string tension scale, the experimental
value for light quarks corresponds to 0.075 in lattice units,
consistent with the three lightest Wilson points and all the SW
points.

\begin{figure}[htbp]
\begin{center}
\leavevmode
\epsfysize=300pt
  \epsfbox[20 30 620 600]{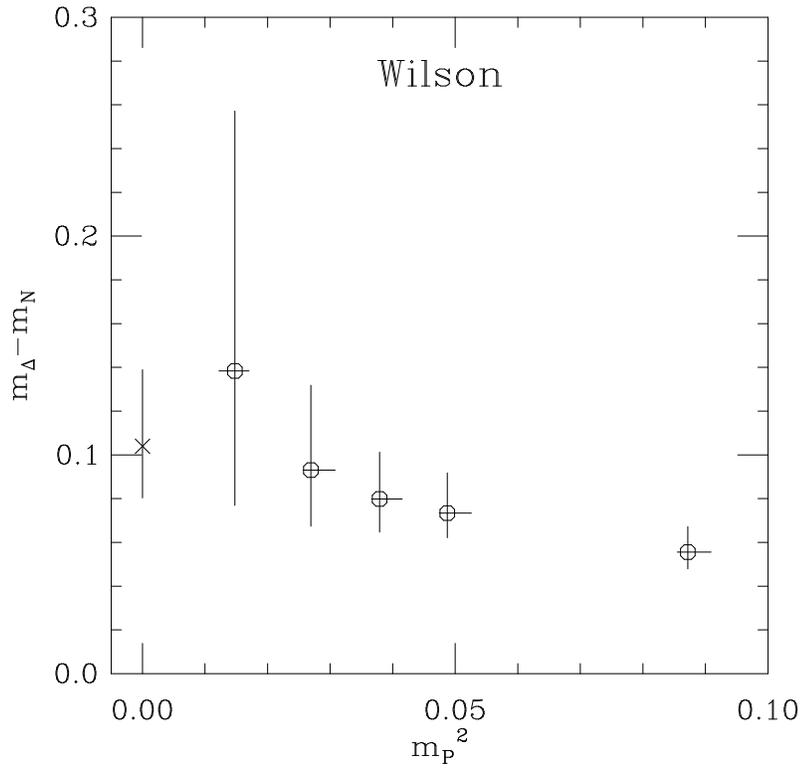}
\end{center}
\begin{center}
\leavevmode
\epsfysize=300pt
  \epsfbox[20 30 620 600]{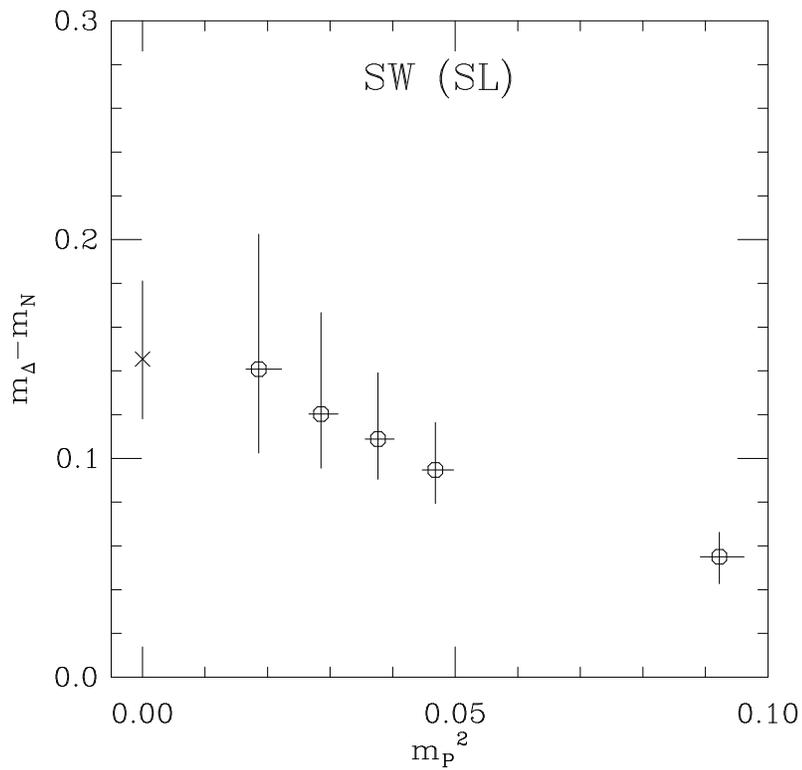}
\end{center}
\caption{$\Delta$-nucleon mass splitting; the
left-most point in each plot is obtained from the chiral
extrapolation of the individual masses.}
\label{delta-nucleon}
\end{figure}

The $\Delta$-nucleon mass splittings for the two actions,
estimated from the bootstrap samples, are shown in
Figure~\ref{delta-nucleon}.
Using the scale set by the string tension, the experimental result
of 300~MeV would translate to 0.11 in lattice units, a value
broadly consistent with both the Wilson and the SW data.
However, the errors in both data sets are too large for this
quantity to discriminate between the two actions.

It has been noted that the ratio
$(m_{K^\ast}-m_\rho)/(m_K^2-m_\pi^2)$ typically gives a value for
the inverse lattice spacing which is lower than that obtained
using other physical quantities~\cite{MandM}.  Assuming that the
vector meson mass is linear in the quark masses, with slope
$b_V$ and intercept $a_V$:
\begin{equation}
m_V(\kappa_1,\kappa_2) = a_V + \frac{b_V}{b_{P}} m_{P}^2(\kappa_1,\kappa_2)
\end{equation}
we obtain the scales given in Table~\ref{tab:scales} from
\begin{equation}
\frac{m_{K^\ast} - m_\rho}{m_K^2 - m_\pi^2} =
\frac{m_V(\kappa_s,\kappa_{crit})   - m_V(\kappa_{crit},\kappa_{crit})}
{m_{P}^2(\kappa_s,\kappa_{crit}) -
m_{P}^2(\kappa_{crit},\kappa_{crit})} = \frac{b_V}{b_{P}}.
\end{equation}

We find that the scales for the two actions are in good agreement with
each other, but our best determinations are more than one standard
deviation below the baryon scales and more than two standard
deviations below the string tension scale.

\section{Meson Decay Constants}

\subsection{Pseudoscalar Decay Constants}

We determine the pseudoscalar decay constant, $f_{P}$, through
the matrix element of the fourth component of the lattice axial current:
\begin{equation}
Z_A\langle0|\bar q(0)\gamma _4\gamma_5q(0)|P(p)\rangle=f_{P}m_{P}
\end{equation}
where our normalisation is such that the physical value of $f_\pi$
is 132 MeV.  The factor $Z_A$ is required to ensure that the
lattice current obeys the correct current algebra in the continuum
limit~\cite{bochicchio}.

Because the signal for $\sum_{\bm{x}} \langle A_4(\bm{x},t)
A_4^\dagger(0) \rangle$ is unacceptably noisy, for both the Wilson
and SW actions, we determine $f_\pseudo$ using the LL
propagators by fitting the ratio
\begin{equation}
\frac{\sum_{\bm{x}} \langle A_4(\bm{x}, t)\,A_4^\dagger(0) \rangle}
{\sum_{\bm{x}} \langle P(\bm{x},t)\,P^\dagger(0)\rangle}
\sim \frac{f_\pseudo^2 m_\pseudo^2}{Z_A^2|\langle 0 | P | \pseudo \rangle|^2}.
\end{equation}
where $m_\pseudo$ and $\langle 0|P|P\rangle$ are obtained from a
separate least-$\chi^2$ fit to $\sum_{\bm{x}} \langle P(\bm{x},t)
P^\dagger(0) \rangle$.  The errors in $f_\pseudo$ are determined
by a bootstrap analysis on the whole procedure.

The measurements of $f_\pseudo$ reported in~\cite{hadrons_lett}
for the SW action displayed much larger errors than those for
the standard Wilson action.  For the SW action, we
computed the correlator $\sum_{\bm{x}} \langle A_4(\bm{x},t)
P^\dagger(0) \rangle$, and thus we are also able to obtain
$f_\pseudo$ by fitting the ratio
\begin{equation}
\frac{\displaystyle \sum_{\bm{x}} \langle A_4(\bm{x},t)
P^{\dagger}(0)
\rangle}
{\displaystyle \sum_{\bm{x}} \langle P(\bm{x},t) P^{\dagger}(0)
\rangle} \sim \frac{f_\pseudo m_\pseudo}{Z_A\langle 0 | P | \pseudo \rangle}
\tanh{m_\pseudo ( L_t/2 - t)},
\end{equation}
where $m_\pseudo$ and $\langle 0|P|P\rangle$ are obtained from
the fit to $\sum_{\bm{x}} \langle P(\bm{x},t) P^\dagger(0) \rangle$.

We found that, for the light hadron sector using Wilson fermions, the
use entirely of local operators provided a good determination of
$f_\pseudo$, and the introduction of SL and SS propagators offered no
improvement ~\cite{smearing}.  Therefore, we did not pursue the
determination of $f_\pi$ using the SL propagators.

In Table~\ref{tab:fpi}, we present the values obtained for $f_{P}$
and for the dimensionless ratio $f_{P}/m_V$, using the Wilson action
with the local axial current, and the SW action with the
`improved' axial current:
\begin{equation}
\bar q(x)(1+\frac{1}{2}\gamma\cdot\dleft )\gamma_\mu\gamma_5
(1-\frac{1}{2}\gamma\cdot\dright )q(x).
\end{equation}
The final row contains results for each column after linear
extrapolation in $1/\kappa$ to the chiral limit.  The measurement of
$f_\pseudo$ for the SW action through the axial-pseudoscalar
correlator is clearly much less noisy than that through the
axial-axial correlator, and will be used in the following discussion.
Our lattice results for $f_{P}/m_V$ vary only slowly with quark mass,
in agreement with the experimental observation that $f_\pi/m_\rho$
(0.17) is approximately the same as $f_K/m_{K^\ast}$ (0.18).  The
chiral extrapolation of this quantity may therefore be more reliable
than that of $f_\pseudo$ alone.
\begin{table}
\centering
\begin{tabular}{|c|c|c|}\hline
\multicolumn{3}{|c|}{Wilson using $\sum_{\bm{x}} \langle
A_4(\bm{x},t) A_4^\dagger(0) \rangle$} \\ \hline
$\kappa$ & $(f_{P})/Z_A^W$    & $(f_{P}/m_V )/Z_A^W$ \\ \hline
0.1510   & 0.081\er{4}{3}  & 0.22\er{1}{1} \\
0.1520   & 0.069\er{5}{5}  & 0.21\er{1}{2} \\
0.1523   & 0.066\er{5}{9}  & 0.21\er{2}{3} \\
0.1526   & 0.065\err{5}{12} & 0.21\er{2}{4} \\
0.1529   & 0.067\err{7}{16} & 0.23\er{3}{6} \\ \hline
0.15328  & 0.056\er{8}{9}  & 0.21\er{2}{4} \\ \hline
\end{tabular}\\[5mm]

\begin{tabular}{|c|c|c||c|c|c|}\hline
\multicolumn{3}{|c||}{SW using $\sum_{\bm{x}} \langle
A_4(\bm{x},t) A_4^\dagger(0) \rangle$} &
\multicolumn{3}{c|}{SW using $\sum_{\bm{x}} \langle
A_4(\bm{x},t) P^\dagger(0) \rangle$} \\ \hline
$\kappa$ & $(f_{P})/Z_A^{C}$    & $(f_{P}/m_V )/Z_A^{C}$ &
$\kappa$ & $(f_{P})/Z_A^{C}$    & $(f_{P}/m_V )/Z_A^{C}$ \\ \hline
0.14144 & 0.060\er{5}{3}  & 0.15\er{1}{1}
 & 0.14144 & 0.064\er{2}{2} & 0.16\er{1}{1} \\
0.14226 & 0.048\er{7}{4}  & 0.14\er{2}{2}
 & 0.14226 & 0.052\er{2}{3} & 0.15\er{1}{1} \\
0.14244 & 0.046\er{8}{6}  & 0.14\er{2}{2}
 & 0.14244 & 0.049\er{2}{3} & 0.15\er{1}{2} \\
0.14262 & 0.046\er{8}{8}  & 0.14\er{3}{3}
 & 0.14262 & 0.047\er{2}{3} & 0.14\er{1}{2} \\
0.14280 & 0.046\err{11}{10}& 0.15\er{5}{3}
 & 0.14280 & 0.044\er{3}{5} & 0.14\er{2}{2} \\ \hline
0.14313 & 0.037\err{12}{8} & 0.13\er{4}{3}
 & 0.14313 & 0.039\er{2}{4} & 0.13\er{1}{2} \\ \hline
\end{tabular}
\caption{Values of the pseudoscalar decay constant, in lattice units, and the
ratio $f_{P}/m_V$.  The last row in each table contains values
obtained by a linear extrapolation to the chiral limit.}
\label{tab:fpi}
\end{table}

In order to determine the physical values, the lattice results given
in Table~\ref{tab:fpi} need to be multiplied by the appropriate
renormalisation constant, $Z_A^W$ or $Z_A^C$.
\begin{equation}
Z_A^W \simeq 1 - 0.132 g^2
\end{equation}
in perturbation theory.  If we use the bare coupling constant as
the expansion parameter, then $Z_A^W\simeq 0.87$ and we find, for
the Wilson action, $f_\pi/m_\rho = 0.18$\er{2}{3}.  However,
reference~\cite{lm} proposes the use of an `effective coupling',
$g_{\rm eff}$, defined through
\begin{equation}
\label{eq:eff_coupling}
g^2_{\rm eff} = \frac{g^2_0}{\langle \frac{1}{3} {\rm Tr} U_{\Box} \rangle}
\Bigl( 1 + \frac{0.513}{4 \pi} g^2_{\rm eff} + O(g^4_{\rm eff})\Bigr).
\end{equation}
where $g_0$ is the bare coupling. At $\beta = 6.2$, $g^2_{\rm eff}
\simeq 1.75 g^2_0$ yielding $Z_A^W\simeq 0.78$, and we obtain
$f_\pi/m_\rho = 0.16$\er{2}{3}.

The perturbative estimate~\cite{pittori} of the renormalisation
constant $Z_A^C$ is close to 1,
\begin{equation}
Z_A^C \simeq 1 - 0.0177 g^2;
\end{equation}
using the bare coupling leads to $Z_A^C\simeq 0.98$, whereas the
effective coupling gives $Z_A^C\simeq 0.97$, yielding, for the
SW action, $f_\pi/m_\rho = 0.13$\er{1}{2} in both cases.  We
note that the uncertainty in $Z_A$ due to the choice of the
perturbative expansion parameter is about 10\% with the Wilson
action and only about 1\% with the SW action.

A recent non-perturbative estimate of $Z_A^C$, based on the use of
chiral Ward identities, gave the result $Z_A^C=1.09(3)$, about 10\%
higher than the one-loop perturbative values quoted
above~\cite{cts_lat92}.  This result was obtained from a simulation at
$\beta=6.0$ using a single value of the quark mass. Although it is
expected that the dependence of $Z_A^C$ on the lattice spacing and
quark mass should be very mild, we feel that this expectation should
be checked before the non-perturbative value is adopted in the present
calculation. We note however that if the non-pertubative value of
$Z_A^C$ proves to be stable, the discrepancy between the result for
$f_\pi/m_\rho$ which we obtain using the SW action and the physical
value is considerably reduced. This underlines the importance of
reliable non-perturbative determinations of the renormalisation
constants in order to get better estimates of the remaining lattice
systematic errors, such as quenching.

\begin{figure}[tbp]
\begin{center}
\leavevmode
\epsfysize=300pt
  \epsfbox[20 30 620 600]{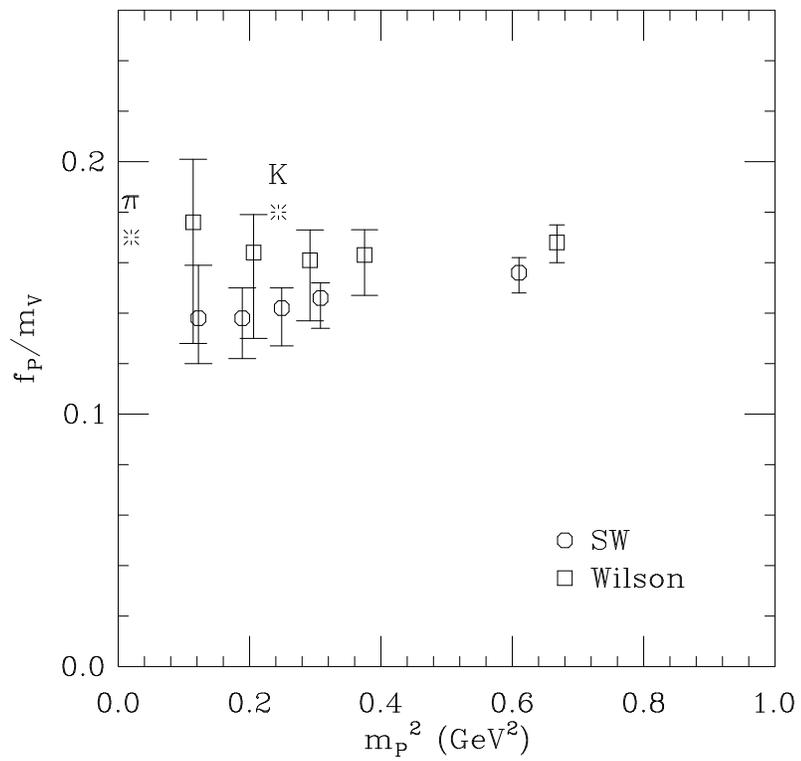}
\end{center}
\caption{$f_{P}/m_V$ against $m^2_{P}$, with lattice spacing
determined from $m_\rho$, and $Z_A$ computed using the effective coupling.}
\label{fig:fps_vs_mps_sq}
\end{figure}

Figure~\ref{fig:fps_vs_mps_sq} shows $f_\pseudo/m_V$ against
$m^2_\pseudo$ in physical units for both actions, with the lattice
spacing determined from $m_\rho$, and $Z_A$ computed using the
effective coupling.  Although the behaviour of this ratio with
$m^2_\pseudo$ for the SW action is very encouraging, there is
a clear discrepancy with the physical values in the overall
normalisation.

If we assume that the pseudoscalar decay constant obeys
\begin{equation}
f_{P} = a_f + b_f m_{P}^2(\kappa_1,\kappa_2),
\end{equation}
where $m_{\pseudo}^2(\kappa_1, \kappa_2)$ is given by
Equation~(\ref{quark_mass}),
then
\begin{equation}
\frac{f_K}{f_\pi}-1 = \frac{b_f}{a_f} m_{P}^2(\kappa_s,\kappa_{crit}).
\end{equation}
We take $a_f$ and $b_f$ from the fit to the data in
Table~\ref{tab:fpi}, and using $\kappa_s$ values from
Equation~(\ref{eq:kappa_s}) we obtain
\begin{equation}
\frac{f_K}{f_\pi}-1 = 0.16\mbox{\err{10}{5}} \quad \mbox{(Wilson)}, \qquad
\frac{f_K}{f_\pi}-1 = 0.25\mbox{\er{7}{4}} \quad \mbox{(SW)},
\end{equation}
compared to the experimental value of 0.22.  Note that the ratio
is less sensitive to uncertainties both in the renormalisation of
the axial current and in the scale.

\subsection{Vector Meson Decay Constant}


The vector meson decay constant is defined by the relation
\begin{equation}
Z_V\langle0|\bar q(0)\gamma _\mu q(0)|V\rangle=\frac{m_V ^2}{f_V}
\epsilon_\mu
\label{eq:frho}
\end{equation}
where $\epsilon_\mu$ is the polarisation vector of the meson and
$Z_V$ is the renormalisation constant for the lattice vector
current~\cite{bochicchio}.  We determine $f_V$, using the LL
propagators, by fitting to
\begin{equation}
\sum_{j = 1}^3 \sum_{\bm{x}} \langle V_j(\bm{x}, t)
V_j^\dagger (0)\rangle \sim \frac{3 m_V^3}{2 Z_V^2 f_V^2}
e^{m_V L_t/2} \cosh m_V (L_t/2 - t).
\end{equation}
For the Wilson action we use the local current and for the SW
action we use the improved local current.  We obtain the values shown
in Table~\ref{tab:frho}.  We extrapolate linearly in $1/\kappa$ to
obtain the values at the chiral limit.
\begin{table}
\centering
\begin{tabular}{|c|c||c|c|}\hline
\multicolumn{2}{|c||}{Wilson} & \multicolumn{2}{c|}{SW} \\ \hline
$\kappa$& $1/(Z_V^Wf_V)$ & $\kappa$ & $1/(Z_V^Cf_V)$ \\ \hline
0.1510  & 0.40\er{1}{1} & 0.14144 & 0.33\er{1}{2} \\
0.1520  & 0.43\er{1}{1} & 0.14226 & 0.36\er{2}{1} \\
0.1523  & 0.44\er{1}{1}  & 0.14244 & 0.36\er{2}{1} \\
0.1526  & 0.45\er{1}{1} & 0.14262 & 0.37\er{2}{1} \\
0.1529  & 0.47\er{2}{1}  & 0.14280 & 0.36\er{2}{1} \\ \hline
0.15328 & 0.47\er{2}{1}  & 0.14313 & 0.38\er{3}{1} \\ \hline
\end{tabular}
\caption{ Values of the vector meson decay constant.  The last row contains
values obtained by a linear extrapolation to the chiral limit.}
\label{tab:frho}
\end{table}

The renormalisation constants corresponding to our choice of the
lattice vector currents are given in perturbation theory
by~\cite{pittori}
\begin{eqnarray}
Z_V^W & \simeq & 1 - 0.17g^2\\
Z_V^C & \simeq & 1 - 0.10g^2.
\end{eqnarray}
The one-loop perturbative correction, although smaller for the
SW action than for the Wilson action, is still substantial and
it introduces a significant uncertainty in $f_V$ due to the
uncertainty in the value of the expansion parameter.  Recent
non-perturbative estimates of $Z_V^C$ at
$\beta=6.0$~\cite{cts_lat92} give a value which agrees well with
that obtained by using the effective coupling defined in
Equation~(\ref{eq:eff_coupling}).  Since the non-perturbative values
of $Z_V$ for the two actions are not yet known at $\beta=6.2$, we
use the effective coupling in the perturbative expressions for
comparison with experiment, i.e.,
\begin{eqnarray}
\label{eq:Z_V}
Z_V^W & \simeq & 0.71\\
Z_V^C & \simeq & 0.83.
\end{eqnarray}
We note that the perturbative uncertainty would be removed
entirely by use of the conserved vector current.

\begin{figure}[tbp]
\begin{center}
\leavevmode
\epsfysize=300pt
  \epsfbox[20 30 620 600]{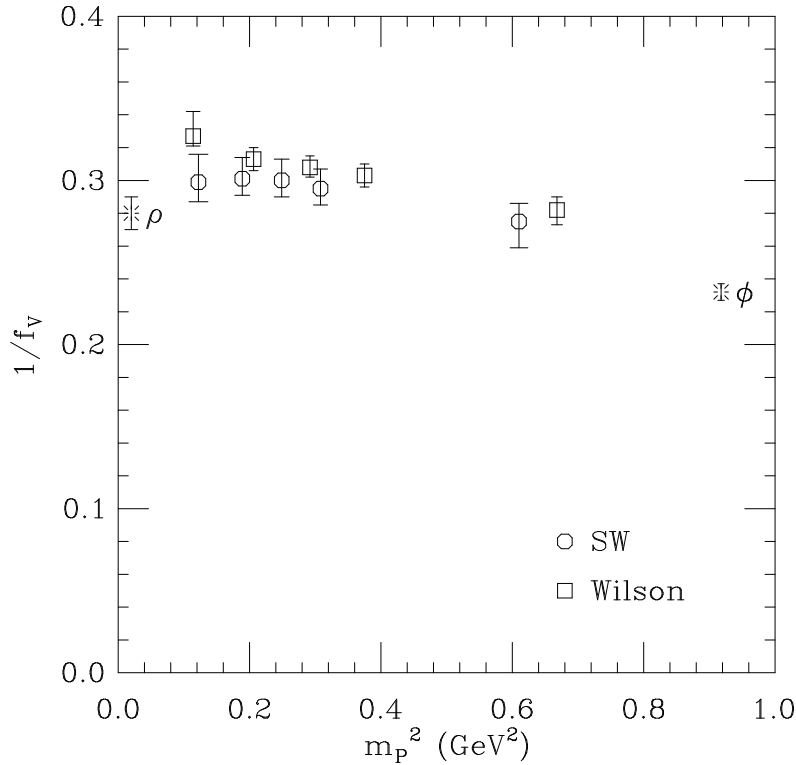}
\end{center}
\caption{$1/f_{V}$ against $m^2_{P}$, with lattice spacing
determined from $m_\rho$, and $Z_V$ computed using the effective
coupling.}
\label{fig:fv_vs_mps_sq}
\end{figure}

Our results for $f_V$ are compared with experimental values in
Figure~\ref{fig:fv_vs_mps_sq}.
We do not know which pseudoscalar meson mass to associate with the
experimental value of $f_\phi$ in this plot.  For the purpose of
illustration, we take the pseudoscalar meson mass to be the mass
of the $\eta'$.  Despite this uncertainty and the presumably small
residual uncertainty in the overall normalisation, we find the
agreement with experiment encouraging.  In the chiral limit we
obtain
\begin{eqnarray}
1/f_\rho & = & 0.33\er{1}{1}\hspace{10mm}{\rm (Wilson)}\\
1/f_\rho & = & 0.31\er{2}{1}\hspace{10mm}{\rm (SW)}
\end{eqnarray}
compared to the physical value of $1/f_\rho$ of 0.28(1).  Again
this discussion points to the need for a non-perturbative
determination of the current renormalisation constants.

\section{Conclusions}

Significant differences have already been reported between the mass
splittings obtained using the SW and Wilson fermion actions for
systems involving heavy quarks~\cite{UKQCD_hyperfine,aida_lat91}.  To
obtain a complete picture of the effects of the $O(a)$-improvement
proposed by Sheikholeslami and Wohlert, it is necessary also to study
the light quark sector.  Our main conclusion is that at $\beta=6.2$,
for pseudoscalar meson masses in the range 330--800~MeV, there is no
statistically significant difference between the results for the
hadron spectrum or decay constants obtained with the two actions.
This supports the validity of results obtained for light hadrons with
the Wilson action over the last few years.

We find that the scales obtained from the meson sector are consistent
with that from the string tension, but there is evidence of
inconsistent scales from the baryon sector and from the ratio
$(m_{K^\ast}-m_\rho)/(m_K^2-m_\pi^2)$.

Our calculation of $f_P/m_V$ with the SW action yields a
result whose dependence on the quark mass is consistent with
experiment, but whose magnitude is significantly smaller.  There
is a small residual uncertainty in the calculated values, which
could be removed by a non-perturbative determination of the axial
vector current renormalisation constant.  It seems likely that
this will not account entirely for the discrepancy and may signal
the effect of quenching. Our results for $f_\rho$  and $f_\phi$
are broadly in agreement with experiment.

We believe that this comparative study has established the
viability of the SW formulation of quenched lattice QCD, with
no significant disadvantages for light hadrons, laying the
foundation for its use in the study of heavy-quark systems where
discretisation errors are more serious.

\subsection*{Acknowledgements}

This research is supported by the UK Science and Engineering Research
Council under grants GR/G~32779, GR/H~49191, GR/H~53624 and
GR/H~01069, by the University of Edinburgh and by Meiko Limited.  The
SERC is acknowledged for its support of CTS through the award of a
Senior Fellowship and of ADS through the award of a Personal
Fellowship.  We are grateful to Edinburgh University Computing Service
for use of its DAP 608 for some of the analysis and, in particular, to
Mike Brown for his tireless efforts in maintaining service on the
Meiko i860 Computing Surface.


\begin{thebibliography}{99}

\bibitem{luscher-weisz}
M.~L\"{u}scher \& P.~Weisz, Commun.~Math.~Phys. 97 (1985) 59.

\bibitem{hmprs} G.~Heatlie, C.T.~Sachrajda, G.~Martinelli, C.~Pittori
\& G.C.~Rossi, Nucl.~Phys. B352 (1991) 266.

\bibitem{symanzik}
K.~Symanzik, {\it in\/} Mathematical Problems in Theoretical Physics,
ed.~R.~Schrader, R.~Seiler \& D.A.~Uhlenbrock,
Springer Lecture Notes in Physics, vol.~153 (1982) 47.

\bibitem{wetzel}
W.~Wetzel, Phys.~Lett.\ B136 (1984) 407.

\bibitem{sheikholeslami}
B.~Sheikholeslami \& R.~Wohlert, Nucl.\ Phys.\ B259 (1985) 572.

\bibitem{hadrons_lett}
UKQCD Collaboration, C.R.~Allton et al., Phys.~Lett.\ B284 (1992) 377.

\bibitem{kcb1}
K.C.~Bowler, Phys.~Rep. {\bf 207} (1991) 261.

\bibitem{kcb2}
K.C.~Bowler, {\em The UK Grand Challenge Projects: Large-Scale
Scientific Computations on a Parallel Supercomputer\/}, Proceedings of
COMETT Seminar on Industrial Applications of Parallel Computers, Graz,
Austria (1992).

\bibitem{chep92}
S.P.~Booth, {\em Parallel Computing in UKQCD\/}, Proceedings of
CHEP92, CERN (1992).

\bibitem{Cr}
M.~Creutz, Phys.~Rev.~D36 (1987) 2394; F.R.~Brown \& T.J.~Woch,
Phys.~Rev.~Lett.~58 (1987) 2394.

\bibitem{C_M}
N.~Cabibbo \& E.~Marinari, Phys.\ Lett.\ B119
(1982) 387.

\bibitem{uni}
G.~Marsaglia, A.~Zaman \& W.W.~Tsang,
Stat.\ Probabil.\ Lett.\ 9 (1990) 35.

\bibitem{T}
M.~Teper, Phys.~Lett.\ B183 (1987) 345.

\bibitem{APE}
M.~Albanese et al., Phys.~Lett.\ B192 (1987) 163.

\bibitem{PM}
S.~Perantonis \& C.~Michael, Nucl.~Phys.~B 347 (1990) 854.

\bibitem{MT}
C.~Michael \& M.~Teper, Nucl.~Phys.~B 314 (1989) 347.

\bibitem{BS}
G.S.~Bali \& K.~Schilling, Phys.\ Rev.\ D46 (1992) 2636.

\bibitem{lm}
G.P.~Lepage \& P.B.~Mackenzie, Nucl.~Phys.~B (Proc.~Suppl.) 20 (1991)
173; {\it On the viability of lattice perturbation theory\/}, Fermilab
preprint FERMILAB-PUB-91-355-T-REV, September 1992.

\bibitem{ukqcd_su3}
UKQCD Collaboration, S.P.~Booth et al., Phys.~Lett.\ B294 (1992) 385.

\bibitem{oyanagi}
Y.~Oyanagi, Comp.~Phys.~Comm. 42 (1986) 333

\bibitem{hockney}
G.~M.~Hockney, Nucl.~Phys. B (Proc. Suppl.) 17 (1990) 301

\bibitem{simpson}
A.D.~Simpson, {\it Algorithms for lattice QCD\/}, Ph.D.~thesis,
University of Edinburgh (1991).

\bibitem{MSV}
G.~Martinelli, C.T.~Sachrajda \& A.~Vladikas, Nucl.~Phys.~B358 (1992)
212.

\bibitem{MSSV}
G.~Martinelli, C.T.~Sachrajda, G.~Salina \& A.~Vladikas,
Nucl.~Phys.~B378 (1992) 591.

\bibitem{Wuppertal}
S.~G\"{u}sken et al., Nucl.~Phys. B (Proc. Suppl.) 17 (1990) 301.

\bibitem{LANL}
D.~Daniel, R.~Gupta, G.W.~Kilcup, A.~Patel and S.~Sharpe, Phys.~Rev.\
D46 (1992) 3130.

\bibitem{kenway84}
R.~D.~Kenway, in Proceedings of XII International Conference on HEP,
Leipzig (1984) 51,  eds.\ A.~Meyer and E.~Wieczorek.

\bibitem{billoire85}
A.~Billoire, E.~Marinari, and G.~Parisi, Phys.\ Lett.\ 162B (1985) 160.

\bibitem{smearing}
UKQCD Collaboration, C.R.~Allton et al., Phys.\ Rev.\ D47 (1993) 5128.

\bibitem{bootstrap}
M.-C.~Chu, M.~Lissia \& J.W.~Negele, Nucl.~Phys. B360 (1991) 31.

\bibitem{APE_hyperfine}
The APE Collaboration: S.~Cabasino et al., Phys.\ Lett.\ B258 (1991) 195.

\bibitem{hyperfine}
C.R.~Allton, M.~Bochiccio, D.B.~Carpenter, G.~Martinelli \&
C.T.~Sachrajda, Nucl.~Phys.~B372 (1992) 403.

\bibitem{UKQCD_hyperfine}
UKQCD Collaboration, C.R.~Allton et al., Phys.~Lett.~B292 (1992) 408.

\bibitem{MandM} L.~Maiani \& G.~Martinelli, Phys.\ Lett.\ B178
(1986) 265.

\bibitem{bochicchio} M.~Bochicchio, L.~Maiani, G.~Martinelli,
G.~Rossi \& M.~Testa, Nucl.~Phys.~B262 (1985) 331.

\bibitem{pittori}
A.~Borrelli, C.~Pittori, R.~Frezzotti \& E.~Gabrielli, {\it New
improved operators: a convenient redefinition\/}, CERN preprint
TH.6587/92 (1992).

\bibitem{cts_lat92} C.T.~Sachrajda, Nucl.\ Phys.\ B (Proc.\ Suppl.) 30
(1993) 20.

\bibitem{aida_lat91} A.X.~El-Khadra, Nucl.\ Phys.\ B
(Proc.\ Suppl.) 26 (1992) 372.

\end{thebibliography}
\end{document}